\documentstyle[12pt]{article}
\input amssym.def
\input amssym.tex

\newcommand{\wa}{\widetilde{a}}
\newcommand{\wv}{\widetilde{v}}

\begin{document}

\renewcommand{\theequation}{\thesection.\arabic{equation}}

\begin{titlepage}{\LARGE
\begin{center} Discrete time Bogoyavlensky lattices \end{center}}

\vspace{1.5cm}

\begin{flushleft}{\large Yuri B. SURIS}\end{flushleft} \vspace{1.0cm}
Centre for Complex Systems and Visualization, Unversity of Bremen,\\
Postfach 330 440, 28334 Bremen, Germany\\
e-mail: suris @ mathematik.uni-bremen.de 

\vspace{2.0cm}

{\small {\bf Abstract.} Discretizations of the Bogoyavlensky lattices are
introduced, belonging to the same hierarchies as the continuous--time systems. 
The construction exemplifies the general scheme for integrable discretization 
of systems on Lie algebras with $r$--matrix Poisson brackets. 
An initial value problem for the difference equations is solved in terms of a 
factorization problem in a group. Interpolating Hamiltonian flow is found.}
\end{titlepage}

\setcounter{equation}{0}
\section{Introduction}
The subject of integrable symplectic maps received in the recent
years a considerable attention.
Given an integrable system of ordinary differential equations with such
attributes as Lax pair, $r$--matrix and so on, one would like to construct
its difference approximation, desirably also with a (discrete--time analog of)
Lax pair, $r$--matrix etc. Recent years brought us several successful
examples of such a construction [1--10].

Recently, stimulated by the results of \cite{Discr CM}, \cite{Discr RS}, 
there was formulated a general recipe for producing  discretizations
sharing the Lax matrix with the continuous--time system, 
so that the discrete--time system belongs to the 
{\it same} integrable hierarchy as the underlying continuous--time one
\cite{New discr Toda}, \cite{Discr RTL}, \cite{Discr peakons}.

In the present paper we want to describe a new application of this scheme to
the Bogoyavlensky lattices \cite{B}, which were given an $r$--matrix 
interpretation in \cite{rmatBog}. Some of equations derived here appeared
previously in the literature \cite{Hirota}, as certain reductions of the
discrete KP equation in the bilinear form. Our approach enables to get these
equations systematically, and, moreover, provides automatically the
Hamiltonian formulation along with the interpolating Hamiltonian flow,
as well as the solution in terms of matrix factorizations.

\setcounter{equation}{0}
\section{Continuous--time Bogoyavlensky lattices}

The Bogoyavlensky lattices were introduced in \cite{B} as three families
of integrable lattice systems depending on integer parameter $m\ge 1$ ($m>1$
for the third one):
\begin{equation}\label{Lat1}
\dot{a}_k=a_k\left(\sum_{j=1}^ma_{k+j}-\sum_{j=1}^ma_{k-j}\right),
\end{equation}
\begin{equation}\label{Lat2}
\dot{a}_k=a_k\left(\prod_{j=1}^ma_{k+j}-\prod_{j=1}^ma_{k-j}\right),
\end{equation}
\begin{equation}\label{Lat3}
\dot{a}_k=a_k\left(\prod_{j=0}^ma_{k+j}^{-1}-
\prod_{j=0}^ma_{k-j}^{-1}\right)
=\prod_{j=1}^ma_{k+j}^{-1}-\prod_{j=1}^ma_{k-j}^{-1}.
\end{equation}

We shall call these systems lattice 1, lattice 2, and lattice 3, respectively.

The lattices 1 and 2 serve as generalizations of the famous Volterra lattice, 
\begin{equation}\label{Vol}
\dot{a}_k=a_k(a_{k+1}-a_{k-1}),
\end{equation}
which is $m=1$ special case of both the systems (\ref{Lat1}), (\ref{Lat2}).
Some special case of the lattice 1 was found also independently by Itoh
\cite{Itoh}.

The lattice 3 after the change of variables $a_k\mapsto a_k^{-1}$ and $t\mapsto 
-t$ turns into
\begin{equation}\label{modLat3}
\dot{a}_k=a_k^2\left(\prod_{j=1}^ma_{k+j}-\prod_{j=1}^ma_{k-j}\right),
\end{equation} 
which serves as a generalization of the so--called modified Volterra lattice, 
the $m=1$ particular case of (\ref{modLat3}):
\begin{equation}\label{modVol}
\dot{a}_k=a_k^2(a_{k+1}-a_{k-1}).
\end{equation}

All these systems may be considered on an infinite lattice (all the subscripts 
belong to ${\Bbb Z}$), and admit also periodic finite--dimensional reductions 
(all the subscripts belong to ${\Bbb Z}/N{\Bbb Z}$, where $N$ is the number of 
particles). The lattices 1 and 2 admit also finite--dimensional versions with 
boundary conditions of the open--end type:
\[
{\rm for\;\;system\;\;}(\ref{Lat1}):
\quad a_k=0\quad{\rm for}\quad k\le 0,\;k\ge N-m+1;
\]
\[
{\rm for\;\;system\;\;}(\ref{Lat2}):
\quad a_k=0\quad{\rm for}\quad k\le 0,\;k\ge N.
\]
Bogoyavlensky has found also the Lax representations for these systems of 
the form
\begin{equation}\label{Lax}
\dot{T}=\left[T,B\right],
\end{equation}
where for the system (\ref{Lat1})
\begin{equation}\label{Tmat1}
T(a,\lambda)=\lambda^{-m}\sum a_kE_{k,k+m}+\lambda\sum E_{k+1,k},
\end{equation}
\begin{equation}\label{Bmat1}
B(a,\lambda)=\sum(a_k+a_{k-1}+\ldots+a_{k-m})E_{k,k}+\lambda^{m+1}\sum 
E_{k+m+1,k},
\end{equation}
for the system (\ref{Lat2})
\begin{equation}\label{Tmat2}
T(a,\lambda)=\lambda^{-1}\sum a_kE_{k,k+1}+\lambda^m\sum E_{k+m,k},
\end{equation}
\begin{equation}\label{Bmat2}
B(a,\lambda)=-\lambda^{-m-1}\sum a_ka_{k+1}\ldots a_{k+m}E_{k,k+m+1},
\end{equation}
and for the system (\ref{Lat3})
\begin{equation}\label{Tmat3}
T(a,\lambda)=\lambda^{-1}\sum a_kE_{k,k+1}+\lambda^{-m-1}\sum E_{k,k+m+1},
\end{equation}
\begin{equation}\label{Bmat3}
B(a,\lambda)=\lambda^m\sum a_k^{-1}a_{k+1}^{-1}\ldots a_{k+m-1}^{-1}E_{k+m,k},
\end{equation}

Here for the infinite lattices all the subscripts belong to ${\Bbb Z}$,
for the periodic cas eall the subscripts belong to ${\Bbb Z}/N{\Bbb Z}$,
and for the open--end case all the subscripts belong to ${1,\ldots,N}$.
Moreover, in the infinite--dimensional and open--end cases the dependence
on the spectral parameter $\lambda$ becomes inessential and may be suppressed
by setting $\lambda=1$. Below we consider only finite lattices.

All the Bogoyavlensky lattices are Hamiltonian systems. More precisely,
each system (\ref{Lat1}), (\ref{Lat2}),  (\ref{Lat3}) is Hamiltonian with
respect to a certain quadratic Poisson bracket
\begin{equation}\label{Dirac}
\{a_k,a_j\}=\pi_{kj}a_ka_j,
\end{equation}
with a skew--symmetric matrix $(\pi_{kj})$.
The corresponding Hamiltonians are:
\[
H(a)={\rm tr}(T^{m+1})/(m+1)=\sum a_k\quad
{\rm for\;\;the\;\;systems\;\;(\ref{Lat1})},
\]
\[
H(a)={\rm tr}(T^{m+1})/(m+1)=\sum a_ka_{k+1}\ldots a_{k+m-1}\quad
{\rm for\;\;the\;\;systems\;\;(\ref{Lat2})},
\] 
\[
H(a)=-{\rm tr}(T^{-m})/m=\sum a_k^{-1}a_{k+1}^{-1}\ldots a_{k+m}^{-1}
\quad {\rm for\;\;the\;\;system\;\;(\ref{Lat3})}.
\] 

The Poisson brackets (\ref{Dirac}), i.e. the matrices $(\pi_{kj})$,
in the context of infinite systems were
found for the lattice 1 in the original papers by Bogoyavlensky \cite{B},
and for the lattices 2 and 3 -- in Ref. \cite{Fuchs}. For the finite lattices,
where some subtleties come out, this was done systematically in \cite{rmatBog}.

\setcounter{equation}{0}
\section{Discrete time Bogoyavlensky lattices}

We present now equations of motion of some difference equations which can
be considered as analogs and approximations to the Bogoyavlensky lattices
for the case of the discrete time. The ''Proposition $k$'' ($k=1,2,3$) deals
with the ''discrete time Bogoyavlensky lattice $k$''. We use tilde to denote 
the time shift, so that, for example, $\wv_k=v_k(t+h)$, if $v_k=v_k(t)$.

{\bf Proposition 1.}  {\it The system of difference equations
\begin{equation}\label{eq in v 1}
\wv_k\prod_{j=1}^m(1+h\wv_{k-j})=v_k\prod_{j=1}^m(1+hv_{k+j})
\end{equation}
admits a Lax representation
\[
\widetilde{T}=L^{-1}TL
\]
with the matrices
\begin{equation}\label{Tmat1 v}
T(v,\lambda)=\lambda^{-m}\sum a_kE_{k,k+m}+\lambda\sum E_{k+1,k},
\end{equation}
\begin{equation}\label{Lfactor1}
L(v,\lambda)=\sum\beta_kE_{k,k}+h\lambda^{m+1}\sum E_{k+m+1,k},
\end{equation}
where}
\begin{equation}\label{thru v 1}
a_k=v_k\prod_{j=1}^m(1+hv_{k-j}),\quad \beta_k=\prod_{j=0}^m(1+hv_{k-j}).
\end{equation}

{\bf Proposition 2.}  {\it The system of difference equations
\begin{equation}\label{eq in v 2}
\wv_k\left(1+h\prod_{j=1}^m\wv_{k-j}\right)=
v_k\left(1+h\prod_{j=1}^mv_{k+j}\right)
\end{equation}
admits a Lax representation
\[
\widetilde{T}=UTU^{-1}
\]
with the matrices
\begin{equation}\label{Tmat2 v}
T(v,\lambda)=\lambda^{-1}\sum a_kE_{k,k+1}+\lambda^m\sum E_{k+m,k},
\end{equation}
\begin{equation}\label{Ufactor2}
U(v,\lambda)=I+h\lambda^{-m-1}\sum \gamma_kE_{k,k+m+1},
\end{equation}
where}
\begin{equation}\label{thru v 2}
a_k=v_k\left(1+h\prod_{j=1}^mv_{k-j}\right),\quad 
\gamma_k=\prod_{j=0}^mv_{k+j}.
\end{equation}

{\bf Proposition 3.}  {\it The system of difference equations
\begin{equation}\label{eq in v 3}
\wv_k\left(1+h\prod_{j=0}^m\wv_{k-j}^{-1}\right)=
v_k\left(1+h\prod_{j=0}^m v_{k+j}^{-1}\right)
\end{equation}
admits a Lax representation
\[
\widetilde{T}=L^{-1}TL
\]
with the Lax matrices
\begin{equation}\label{Tmat3 v}
T(v,\lambda)=\lambda^{-1}\sum a_kE_{k,k+1}+\lambda^{-m-1}\sum E_{k,k+m+1},
\end{equation}
\begin{equation}\label{Lfactor3}
L(v,\lambda)=I+h\lambda^m\sum \alpha_kE_{k+m,k},
\end{equation}
where}
\begin{equation}\label{thru v 3}
a_k=v_k\left(1+h\prod_{j=0}^m v_{k-j}^{-1}\right),\quad 
\alpha_k=\prod_{j=0}^{m-1}v_{k+j}^{-1}.
\end{equation}

{\bf Remark 1.} Upon change of variables $v_k\mapsto v_k^{-1}$ and 
$h\mapsto -h$ the system (\ref{eq in v 3}) turns into
\begin{equation}\label{mod eq in v 3 }
\wv_k\left(1-h\prod_{j=0}^m\wv_{k-j}\right)^{-1}=
v_k\left(1-h\prod_{j=0}^m v_{k+j}\right)^{-1},
\end{equation}
which  may be considered as a discrete time analog and approximation
to (\ref{modLat3}).

{\bf Remark 2.} 
The equation (\ref{eq in v 1}) was found in \cite{Hirota}
as a certain reduction of the discrete KP equation in the bilinear form.
Other equations (\ref{eq in v 2}), (\ref{eq in v 3}) seem to be new.
The equations (\ref{eq in v 1}) and (\ref{eq in v 2}) for $m=1$ coincide, 
as they should (Volterra lattice). The Lax representation for this case with
the matrices (\ref{Tmat2 v}), (\ref{Ufactor2}) was also given in \cite{Hirota}, 
but without any hint on how it was obtained.

In the above formulation these Propositionss may be easily checked by a direct
computation, but their origin remains hidden. In the following sections
we shall give a way to {\it derive} them systematically, which, as a by--product,
will unvail an underlying invariant Poisson structure of these discrete
systems, as well as a role of the auxiliary matrices $L,U$. This, in turn,
will enable us to solve the initial value problems for our systems
in terms of matrix factorizations and to find interpolating Hamiltonian flows.
Our construction is just a particular case of a general one, applicable, in
principle, to every system admitting an $r$--matrix interpretation. The key
observation is that the Lax matrices (\ref{Tmat1 v}), (\ref{Tmat2 v}), 
(\ref{Tmat3 v}) of the discrete time systems formally coincide with the 
corresponding Lax matrices (\ref{Tmat1}), (\ref{Tmat2}), (\ref{Tmat3}) of
the continuous time ones.

\setcounter{equation}{0}
\section{Algebraic structure \newline of Bogoyavlensky lattices}

In \cite{rmatBog} we gave an $r$--matrix interpretation of the Bogoyavlensky
lattices as simplest representatives of integrable hierarchies on associative 
algebras. The main results of \cite{rmatBog} may be summarized as follows.

1) For the {\it open--end case} (applies only to the lattices 1 and 2)
we set ${\rm\bf g}=gl(N)$. To this algebra there corresponds a group
${\rm\bf G}=GL(N)$. As a linear space, ${\rm\bf g}$ may be represented 
as a direct sum of two subspaces, which serve also as subalgebras: 
${\rm\bf g}={\rm\bf g}_+\oplus{\rm\bf g}_-$.
Here ${\rm\bf g}_+$ (${\rm\bf g}_-$) is a space of all 
lower triangular (resp. strictly upper triangular)  $N$ by $N$ matrices. 
The corresponding subgroups: ${\rm\bf G}_+$ (${\rm\bf G}_-$) 
is a group of all nondegenerate lower triangular $N$ by $N$ matrices 
(resp. upper triangular $N$ by $N$ matrices with unities on the diagonal).  

2) For the {\it periodic case} (of all lattices 1, 2, 3) ${\rm\bf g}$ is 
a certain twisted loop algebra over $gl(N)$, namely the algebra of formal 
semi--infinite Laurent series $T(\lambda)$ over $gl(N)$, satisfying
$\Omega T(\lambda)\Omega^{-1}=T(\omega\lambda)$, where $\Omega={\rm diag}
(1,\omega,\ldots,\omega^{N-1})$, $\omega=\exp(2\pi i/N)$. The 
corresponding group is the twisted loop group ${\rm\bf G}$ consisting
of $GL(N)$--valued functions $T(\lambda)$ of the complex parameter $\lambda$, 
regular in ${\Bbb C}P^1\backslash\{0,\infty\}$ and satisfying $\Omega T(\lambda)
\Omega^{-1}=T(\omega\lambda)$. Again, as a linear space ${\rm\bf g}=
{\rm\bf g}_+\oplus{\rm\bf g}_-$, where for the lattices 1 and 2 
${\rm\bf g}_+$ (${\rm\bf g}_-$) is a subspace and subalgebra consisting of 
$T(\lambda)$ containing only non--negative (resp. only negative) powers of 
$\lambda$, and the case of the lattice 3 differs in that to which subalgebra
do diagonal matrices belong: ${\rm\bf g}_+$ contains only positive, and
${\rm\bf g}_-$ only non--positive powers of $\lambda$. 
For the lattices 1 and 2 the corresponding subgroups ${\rm\bf G}_+$ and
${\rm\bf G}_-$ consist of $T(\lambda)$ regular in the 
neighbourhood of $\lambda=0$ (resp. regular in the neighbourhood of 
$\lambda=\infty$ and taking the value $I$ in $\lambda=\infty$). For the 
lattice 3 ${\rm\bf G}_+$ is formed by $T(\lambda)$ regular in the 
neighbourhood of $\lambda=0$ with $T(0)=I$, and ${\rm\bf G}_-$ is formed by
$T(\lambda)$ regular in the neighbourhood of $\lambda=\infty$.

For both the open--end and periodic cases every $T\in{\rm\bf g}$ admits
a unique decomposition $T=l(T)+u(T)$, where $l(T)\in{\rm\bf g}_+$, $u(T)\in
{\rm\bf g}_-$. Analogously, for the both cases every $T\in{\rm\bf G}$ from some 
neighbourhood of the group unity admits a unique factorization 
$T={\cal L}(T)\,{\cal U}(T)$, where ${\cal L}(T)\in{\rm\bf G}_+$, 
${\cal U}(T)\in{\rm\bf G}_-$.

There hold the following statements.

{\it 
{\rm a)} For each system {\rm (\ref{Lat1}), (\ref{Lat2}), (\ref{Lat3})}
there exists a quadratic $r$--matrix Poisson bracket on ${\rm\bf g}$ whose 
Dirac reduction to the corresponding set of  matrices 
${\cal P}=\{T(a,\lambda)\}$ from {\rm (\ref{Tmat1}), (\ref{Tmat2})}, or 
{\rm (\ref{Tmat3})}, respectively, is given by {\rm (\ref{Dirac})}.

{\rm b)} Let $\varphi: {\rm\bf g}\mapsto{\Bbb C}$ be an invariant function,
so that 
\footnote{Recall that the gradient $\nabla\varphi(T)\in{\rm\bf g}$ of the 
function $\varphi: {\rm\bf g}\mapsto{\Bbb R}$ is defined in the open--end
case by the relation
\[
{\rm tr}(\nabla\varphi(T)X)=
\left.\frac{d}{d\varepsilon}\varphi(T+\varepsilon X)\right|_{\varepsilon=0}
\quad\forall X \in {\rm\bf g};
\]
in the periodic case ''${\rm tr}$'' should be replaced by ''${\rm tr}_0$'',
the free term in the Laurent series for the trace.}
$d\varphi(T)=T\nabla\varphi(T)=\nabla\varphi(T)T$  is covariant under 
conjugation.
Then the Hamiltonian flow on  ${\rm\bf g}$ with the Hamiltonian function
$\varphi(T^p)/p$ (here and below $p=m+1$ for the lattices 1,2, and $p=m$
for the lattice 3) is tangent to ${\cal P}$ and has the Lax form
\begin{equation}\label{genLax}
\dot{T}=\left[T,l(d\varphi(T^p))\right]=-\left[T,u(d\varphi(T^p))
\right].
\end{equation}
This flow admits the following solution in terms of the factorization
problem
\[
e^{{\displaystyle td\varphi(T^p(0))}}=
{\cal L}(t)\,{\cal U}(t),\quad 
{\cal L}(t)\in {\rm\bf G}_+,
\quad {\cal U}(t)\in {\rm\bf G}_-
\]
(this problem has solutions at least for sufficiently small t):
\[
T(t)={\cal L}^{-1}(t)T(0){\cal L}(t)={\cal U}(t)T(0)\,{\cal U}^{-1}(t).
\]

{\rm c)} Let $f:{\rm\bf g}\mapsto{\rm\bf G}$ be a conjugation covariant 
function on ${\rm\bf g}$. Then the  difference equation 
\begin{equation}\label{dLax}
\widetilde{T}=
{\cal L}^{-1}\Big(f(T^p)\Big)T\,{\cal L}\Big(f(T^p)\Big)
={\cal U}\Big(f(T^p)\Big)T\,{\cal U}^{-1}\Big(f(T^p)\Big)
\end{equation}
defines a Poisson map ${\rm\bf g}\mapsto{\rm\bf g}$ which leaves ${\cal P}$
invariant, the restriction of this map on ${\cal P}$ being Poisson with
respect to the reduced bracket {\rm (\ref{Dirac})}. This difference 
equation admits following solution in terms of the factorization problem
\[
f^n(T^p(0))=
{\cal L}(nh)\,{\cal U}(nh), 
\quad {\cal L}(nh)\in {\rm\bf G}_+,
\quad {\cal U}(nh)\in {\rm\bf G}_-
\]
(this problem has solutions for a given $n$ at least if $f(T(0))$ is 
sufficiently close to the group unity $I$):
\[
T(nh)={\cal L}^{-1}(nh)T(0){\cal L}(nh)=
{\cal U}(nh)T(0)\,{\cal U}^{-1}(nh).
\]

{\rm d)} The solutions of the difference equation {\rm (\ref{dLax})} are
interpolated by the flow {\rm (\ref{genLax})} with the Hamiltonian function
$\varphi(T^p)/p$, where $\varphi(T)$ is defined by
\begin{equation}
d\varphi(T)=h^{-1}\log(f(T)).
\end{equation}
}

The statements a),b) explain the Lax equation (\ref{Lax}) with the matrices 
(\ref{Tmat1})--(\ref{Bmat3}), as for the system (\ref{Lat1}) we have 
$B(a,\lambda)=l(T^{m+1}(a,\lambda))$, for the system (\ref{Lat2})
we have $B(a,\lambda)=-u(T^{m+1}(a,\lambda))$, and for the system
(\ref{Lat3}) we have $B(a,\lambda)=l(T^{-m}(a,\lambda))$.

\setcounter{equation}{0}
\section{A discretization of the 
\newline Bogoyavlensky lattice 1}

We get a correct perspective for the interpretation of the system
(\ref{eq in v 1}) (as well as the systems (\ref{eq in v 2}), (\ref{eq in v 3}))
if we take an ''inverse'' view--point. We consider the first equation in
(\ref{thru v 1}) as an implicit definition of the functions $v_k=v_k(a)$,
rather then the expressions of $a_k$ through $v_j$.
In the open--end case the sequence of $v_k$'s can be computed even explicitly,
term by term, starting with $v_k=a_k/(1+h\sum_{j=1}^{k-1}a_j)$
for $1\le k\le m+1$. In particular, for $m=1$ one has $v_k=a_k/(1+hv_{k-1})$,
which implies a nice representation in form of a finite continued fraction:
\begin{equation}\label{cont fr}
v_k=\frac{a_k}{1+
\displaystyle\frac{ha_{k-1}}{1+
\parbox[t]{0.9cm}{$\begin{array}{c}\\  \ddots\end{array}$}
\parbox[t]{2.2cm}{$\begin{array}{c}
 \\  \\+\displaystyle\frac{ha_2}{1+ha_1}\end{array}$}}}
\end{equation}

In the periodic case the existence of the functions $v_k=v_k(a)$,
at least for $h$ small enough, follows from the implicit functions theorem.
Again, for $m=1$ we get an expression in the form of an infinite $N$--periodic
continued fraction of the type (\ref{cont fr}).

The second equation in (\ref{thru v 1}) may be rewritten as a recurrent 
relation for $\beta_k=\beta_k(a)$.
In fact, we have $\beta_k-ha_k=\prod_{j=1}^m(1+hv_{k-j})$, so that 
$a_k/(\beta_k-ha_k)=v_k$, and finally
\begin{equation}\label{recur beta}
\beta_k-ha_k=\prod_{j=1}^m
\left(1+\frac{ha_{k-j}}{\beta_{k-j}-ha_{k-j}}\right).
\end{equation}
Conversely, the last formula implies (\ref{thru v 1}), if one sets $v_k=
a_k/(\beta_k-ha_k)$.

The formula (\ref{recur beta}) may also serve for a successive computation of 
$\beta_k$'s in the open--end case, and in the periodic case it uniquely defines 
a set of $\beta_k-ha_k$, $1\le k\le N$, via the implicit functions theorem. 
In both cases it is easy to see that
\begin{equation}\label{asymp beta}
\beta_k=1+h\sum_{j=0}^m a_{k-j}+O(h^2).
\end{equation}

{\bf Theorem 1.}  {\it The quantities $\beta_k$ defined by 
{\rm (\ref{recur beta})}
serve as coefficients of the matrix
\begin{equation}\label{Pi+ 1}
L={\cal L}(I+hT^{m+1})=\sum_k\beta_kE_{k,k}+h\lambda^{m+1}\sum_kE_{k+m+1,k}.
\end{equation}
The discrete time Lax equation 
\begin{equation}\label{discrLax 1}
\widetilde{T}=L^{-1}TL={\cal L}^{-1}(I+hT^{m+1})\;T\;{\cal L}(I+hT^{m+1})
\end{equation}
with the Lax matrix {\rm (\ref{Tmat1})} 
generates the following map on ${\Bbb R}^N\{a\}$, equivalent to 
{\rm (\ref{eq in v 1})}:
\begin{equation}\label{ev 1}
\wa_k=\frac{\beta_{k+m}}{\beta_k}\;a_k.
\end{equation}
This map is Poisson with respect to the Poisson bracket {\rm (\ref{Dirac})}
corresponding to the lattice 1, and is interpolated by the flow with the
Hamiltonian function}
\begin{equation}\label{interp}
\frac{1}{m+1}\,{\rm tr}\,\Phi(T^{m+1}),\;\;where\;\;
\Phi(\xi)=h^{-1}\int_0^{\xi}\log(1+h\eta)\frac{d\eta}{\eta}.
\end{equation}

{\bf Proof.} The last two statements follow from the results formulated in the
previous section, provided the first two statements are proved.
Suppose for a moment that the ${\cal L}$--factor of $I+hT^{m+1}$ has the form 
(\ref{Pi+ 1}). Then the evolution equation (\ref{discrLax 1}), i.e. 
$L\widetilde{T}=TL$, is equivalent to:
\begin{equation}\label{aux 1}
\beta_k\wa_k=a_k\beta_{k+m}, \quad 
h\wa_k+\beta_{k+m+1}=ha_{k+m+1}+\beta_{k+m}.
\end{equation}
This in turn is equivalent to a combination of an evolution equation 
(\ref{ev 1}) with the condition of compatibility of two equations in 
(\ref{aux 1}):
\begin{equation}\label{comp 1}
\beta_k-ha_k=\frac{\beta_k}{\beta_{k+m}}(\beta_{k+m+1}-ha_{k+m+1}).
\end{equation}
The last equation is equivalent to the fact that
\begin{equation}\label{const 1}
\frac{\displaystyle\prod_{j=0}^m(\beta_{k+j}-ha_{k+j})}
{\displaystyle\prod_{j=0}^{m-1}\beta_{k+j}}={\rm const},
\end{equation}
i.e. does not depend on $k$. We shall prove that the actual value of this 
constant is equal to 1, which is just equivalent to (\ref{recur beta}).

The inspection of the structure of the matrix $T^{m+1}$ for 
$T$ from (\ref{Tmat1}) convinces that the ${\cal L}$--factor 
of $I+hT^{m+1}$ has in fact the form (\ref{Pi+ 1}), while the ${\cal U}$--factor
has the form 
\[
U={\cal U}(I+hT^{m+1})=
I+h\sum_{j=1}^m\lambda^{-j(m+1)}\sum_k\gamma_k^{(j)}E_{k,k+j(m+1)}.
\]
The quantities $\beta_k$, $\gamma_k^{(j)}$ are completely defined 
by the set of recurrent relations following from the definitions:
\begin{equation}\label{recur 1}
\beta_k+h^2\gamma_{k-m-1}^{(1)}=1+h\sum_{j=0}^ma_{k-j},
\end{equation}
\[
\beta_k\gamma_k^{(j)}+h\gamma_{k-m-1}^{(j+1)}={\rm coef.\;\;by}\;\;
\lambda^{-j(m+1)}E_{k,k+j(m+1)}\;\;{\rm in}\;\;T^{m+1},\;\; 1\le j\le m-1;
\]
\[
\beta_k\gamma_k^{(m)}=\prod_{j=0}^ma_{k+jm}.
\]

Now we are in a position to prove that the constant in (\ref{const 1})
is equal to 1.

Indeed, in the open--end case it is enough to compute from
(\ref{recur 1}) the first $m+1$ values of $\beta_k$, namely 
$\beta_k=1+h\sum_{j=1}^ka_j$, $1\le k\le m+1$, which implies 
$\prod_{j=1}^{m+1}(\beta_j-ha_j)/\prod_{j=1}^m\beta_j=1$.

In the periodic case we have found only a combinatoric proof based on tedious
computations. For the sake of simplicity and in order to avoid complicated 
notations we present the corresponding argument only in the simplest cases 
$m=1,2$.

In the case $m=1$ the defining recurrent relations take the form:
\[
\beta_k+h^2\gamma_{k-2}^{(1)}=1+ha_k+ha_{k-1},\quad
\beta_k\gamma_k^{(1)}=a_ka_{k+1}.
\]
Excluding $\gamma_k^{(1)}$ from these relations, we get:
\[
1=\beta_{k+2}-ha_{k+2}-ha_{k+1}\frac{\beta_k-ha_k}{\beta_k}.
\]
Replacing the fraction on the right--hand side through its expression following 
from (\ref{comp 1}) for $m=1$, we get:
\[
1=\frac{(\beta_{k+2}-ha_{k+2})(\beta_{k+1}-ha_{k+1})}{\beta_{k+1}},
\]
which proves the theorem in the case $m=1$.

For $m=2$ the defining recurrent relations take the form

\[
\beta_k+h^2\gamma_{k-3}^{(1)}=1+ha_k+ha_{k-1}+ha_{k-2},
\]
\[
\beta_k\gamma_k^{(1)}+h\gamma_{k-3}^{(2)}=
a_ka_{k+2}+a_ka_{k+1}+a_{k-1}a_{k+1},\quad
\beta_k\gamma_k^{(2)}=a_ka_{k+2}a_{k+4}.
\]
Excluding from these relations $\gamma_k^{(j)}$, we get:
\[
1=\beta_{k+3}-ha_{k+3}-h(a_{k+2}+a_{k+1})\frac{\beta_k-ha_k}{\beta_k}
+h^2a_{k+1}a_{k-1}\frac{\beta_{k-3}-ha_{k-3}}{\beta_k\beta_{k-3}}.
\]
According to (\ref{comp 1}) for $m=2$, this is equivalent to
\[
1=\beta_{k+3}-ha_{k+3}-
h(a_{k+2}+a_{k+1})\frac{\beta_{k+3}-ha_{k+3}}{\beta_{k+2}}
+h^2a_{k+1}a_{k-1}\frac{\beta_{k+3}-ha_{k+3}}{\beta_{k+2}\beta_{k-1}}.
\]
\[
=\frac{(\beta_{k+3}-ha_{k+3})(\beta_{k+2}-ha_{k+2})}{\beta_{k+2}}-
ha_{k+1}\frac{(\beta_{k+3}-ha_{k+3})(\beta_{k-1}-ha_{k-1})}
{\beta_{k+2}\beta_{k-1}}
\]
Using in the last term once more (\ref{comp 1}) for $m=2$, we obtain
\[
1=\frac{(\beta_{k+3}-ha_{k+3})(\beta_{k+2}-ha_{k+2})(\beta_{k+1}-ha_{k+1})}
{\beta_{k+2}\beta_{k+1}},
\]
which proves the theorem for $m=2$. The pattern of the proof for a general 
$m$ may be seen from these two particular cases.

\setcounter{equation}{0}
\section{A discretization of the 
\newline Bogoyavlensky lattice 2}

For the lattice 2 we again consider the first equation in (\ref{thru v 2})
as a definition of the functions $v_k=v_k(a)$. In the open--end case we
can compute these functions succesively, starting with $v_k=a_k
(1+h\sum_{j=1}^{k-2}\prod_{l=1}^ja_l)/(1+h\sum_{j=1}^{k-1}\prod_{l=1}^ja_l)$ 
for $1\le k\le m+1$. In the periodic case the implicit functions theorem
has to be invoked. In particular, for the case $m=1$ we obtain the same 
continued fractions expressions as in the previous section.

The second equation in (\ref{thru v 2}) may be represented as a recurrent
relation for $\gamma_k=\gamma_k(a)$. Indeed, we have $a_k-h\gamma_{k-m}=v_k$,
so that
\begin{equation}\label{recur gamma}
a_k-h\gamma_{k-m}=\frac{a_k}
{1+h\displaystyle\prod_{j=1}^m(a_{k-j}-h\gamma_{k-m-j})}.
\end{equation}
Conversely, the last formula implies (\ref{thru v 2}), if one sets $v_k=
a_k-h\gamma_{k-m}$.

In the open--end case the formula (\ref{recur gamma}) serves as a basis for 
successive computation of $\gamma_k$'s, and in the periodic case it uniquely
defines, by the implicit function theorem, the quantities $a_{k+m}-h\gamma_k$,
$1\le k\le N$. In both cases there holds the following asymptotic relation:
\begin{equation}\label{asymp gamma}
\gamma_k=\prod_{j=0}^ma_{k+j}(1+O(h)).
\end{equation}

{\bf Theorem 2.}  {\it The quantities $\gamma_k$ defined by 
{\rm (\ref{recur gamma})} serve as coefficients of the matrix
\begin{equation}\label{Pi- 2}
U={\cal U}(I+hT^{m+1})=I+h\lambda^{-(m+1)}\sum_k\gamma_kE_{k,k+m+1}.
\end{equation}
The discrete time Lax equation
\begin{equation}\label{discrLax 2}
\widetilde{T}=UTU^{-1}={\cal U}(I+hT^{m+1})\;T\;{\cal U}^{-1}(I+hT^{m+1})
\end{equation}
with the Lax matrix {\rm (\ref{Tmat2})} 
generates the following map on ${\Bbb R}^N\{a\}$, equivalent to 
{\rm (\ref{eq in v 2})}:
\begin{equation}\label{ev 2}
\wa_k=\frac{a_k-h\gamma_{k-m}}{a_{k+m+1}-h\gamma_{k+1}}\;a_{k+m+1}.
\end{equation}
This map is Poisson with respect to the Poisson bracket {\rm (\ref{Dirac})}
corresponding to the lattice 2, and is interpolated by the flow with the
Hamiltonian function} (\ref{interp}).

{\bf Proof.} Again, it suffices to prove the first two statements.
Assuming for a moment that the ${\cal U}$--factor of the the matrix 
$I+hT^{m+1}$ for $T$ from (\ref{Tmat2}) has the form (\ref{Pi- 2}),
we see that the evolution equation (\ref{discrLax 2}), i.e. 
$\widetilde{T}U=UT$, is equivalent to
\begin{equation}\label{aux 2}
\wa_k\gamma_{k+1}=\gamma_ka_{k+m+1}, \quad \wa_k+h\gamma_{k-m}=a_k+h\gamma_k.
\end{equation}
This in turn is equivalent to a combination of an evolution equation
(\ref{ev 2})
with the condition of compatibility of two equations in (\ref{aux 2}):
\begin{equation}\label{comp 2}
a_k-h\gamma_{k-m}=\frac{\gamma_k}{\gamma_{k+1}}(a_{k+m+1}-h\gamma_{k+1}).
\end{equation}
The last equation is equivalent to the fact that
\begin{equation}\label{const 2}
\frac{1}{\gamma_k}\prod_{j=0}^m(a_{k+m-j}-h\gamma_{k-j})={\rm const},
\end{equation}
i.e. does not depend on $k$. We shall prove that the actual value of 
this constant is equal to 1, which is equivalent to (\ref{recur gamma}).

This time the inspection convinces that the ${\cal U}$--factor of the matrix 
$I+hT^{m+1}$ for $T$ from (\ref{Tmat2}) must indeed have the form
(\ref{Pi- 2}), while the ${\cal L}$--factor must have the form
\[
L={\cal L}(I+hT^{m+1})=\sum_k\beta_k^{(0)}E_{k,k}
+h\sum_{j=1}^{m}\lambda^{j(m+1)}\sum_k\beta_k^{(j)}E_{k+j(m+1),k}
\]
where $\beta_k^{(m)}=1$, and other quantities  $\gamma_k$, $\beta_k^{(j)}$ 
are completely defined by the recurrent relations following from the 
definitions:
\begin{eqnarray}\label{recur 21}
\beta_k^{(0)}\gamma_k&=&\prod_{j=0}^ma_{k+j},\\
\label{recur 22}
\beta_k^{(0)}+h^2\beta_{k-m-1}^{(1)}\gamma_{k-m-1}
&=&1+h\sum_{l=k-m}^k\prod_{j=0}^{m-1}a_{l+j},
\end{eqnarray}
\[
\beta_k^{(j)}+h\beta_{k-m-1}^{(j+1)}\gamma_{k-m-1}=
{\rm coef.\;\;by}\;\;
\lambda^{j(m+1)}E_{k+j(m+1),k}\;\;{\rm in}\;\;T^{m+1},\;\; 1\le j\le m-1.
\]

To prove that the constant in (\ref{const 2}) is equal to 1, in the open--end 
case is enough to compute from (\ref{recur 21}), (\ref{recur 22}) the first
$m+1$ values of $\gamma_k$, namely $\gamma_k=\prod_{j=0}^ma_{k+j}/
\left(1+h\sum_{l=1}^k\prod_{j=0}^{m-1}a_{l+j}\right)$, $1\le k\le m+1$,
which implies $\prod_{j=1}^{m+1}(a_{j+m}-h\gamma_j)/\gamma_{m+1}=1$.

In the periodic case we shall again give the proof only for $m=1,2$, leaving
the tedious calculations for the general case to the reader. For $m=1$ the 
defining recurrences (\ref{recur 21}), (\ref{recur 22}) take the form:
\[
\beta_k^{(0)}\gamma_k=a_ka_{k+1},\quad
\beta_k^{(0)}+h^2\gamma_{k-2}
=1+ha_k+ha_{k-1}.
\]
Excluding from these relations $\beta_k^{(0)}$, we get:
\[
1=\frac{a_k}{\gamma_k}(a_{k+1}-h\gamma_k)-h(a_{k-1}-h\gamma_{k-2}).
\]
Replacing the last term on the right--hand side through its expression 
following from  (\ref{comp 2}) for $m=1$, we get:
\[
1=\frac{(a_{k+1}-h\gamma_k)(a_k-h\gamma_{k-1})}{\gamma_k},
\]
which proves the theorem for $m=1$.

In the case $m=2$ the recurrent relations (\ref{recur 21}), (\ref{recur 22})
take the form
\[
\beta_k^{(0)}\gamma_k=a_ka_{k+1}a_{k+2},\quad
\beta_k^{(0)}+h^2\beta_{k-3}^{(1)}\gamma_{k-3}
=1+h(a_{k-2}a_{k-1}+a_{k-1}a_{k}+a_{k}a_{k+1}),
\]
\[
\beta_k^{(1)}+h\gamma_{k-3}=a_{k-1}+a_{k+1}+a_{k+3}.
\]
Excluding $\beta_k^{(j)}$ from these relations, we get:
\[
1=\frac{a_ka_{k+1}}{\gamma_k}(a_{k+2}-h\gamma_k)
-h(a_{k-2}+a_k)(a_{k-1}-h\gamma_{k-3})
+h^2\gamma_{k-3}(a_{k-4}-h\gamma_{k-6})
\]
Using on the right--hand side repeatedly (\ref{comp 2}) for $m=2$, 
we can rewrite it as
\[
1=\frac{a_ka_{k+1}}{\gamma_k}(a_{k+2}-h\gamma_k)-
\frac{h(a_{k-2}+a_k)\gamma_{k-1}}{\gamma_k}(a_{k+2}-h\gamma_k)
+\frac{h^2\gamma_{k-4}\gamma_{k-1}}{\gamma_k}(a_{k+2}-h\gamma_k)
\]
\[
=\frac{a_k}{\gamma_k}(a_{k+2}-h\gamma_k)(a_{k+1}-h\gamma_{k-1})
-\frac{h\gamma_{k-1}}{\gamma_k}(a_{k+2}-h\gamma_k)(a_{k-2}-h\gamma_{k-4}).
\]
Using in the last term once more (\ref{comp 2}) for $m=2$, we get
\[
1=\frac{1}{\gamma_k}
(a_{k+2}-h\gamma_k)(a_{k+1}-h\gamma_{k-1})(a_k-h\gamma_{k-2}),
\]
which finishes the proof for $m=2$.

\setcounter{equation}{0}
\section{A discretization of the 
\newline Bogoyavlensky lattice 3}

For the lattice 3 we again define the functions $v_k=v_k(a)$ by means of the
first equation in (\ref{thru v 3}), which is justified by the implicit function
theorem (as opposed to the lattices 1, 2, this time an open--end reduction is 
not admissible, so that only the periodic case needs to be considered).
In particular, for $m=1$ we have $v_k=a_k-h/v_{k-1}$, which leads to the
expression in terms of an infinite $N$--periodic continued fraction:
\[
v_k=a_k-\frac{h}{\parbox[t]{1.1cm}{$a_{k-1}-$}
\parbox[t]{0.9cm}{$\begin{array}{c}\\ \ddots\end{array}$}
\parbox[t]{3.1cm}{$\begin{array}{c}\\ \\
-\displaystyle\frac{h}{a_{k-N+1}-
\displaystyle\frac{h}{v_k}}\end{array}$}}
\]

The second equation in (\ref{thru v 2}) implies $a_k-h\alpha_{k-m}=v_k$, and
hence
\begin{equation}\label{recur alfa}
\alpha_k=\prod_{j=0}^{m-1}\frac{1}{a_{k+j}-h\alpha_{k+j-m}}.
\end{equation}
Conversely, the last formula implies (\ref{thru v 3}), if one defines $v_k=
a_k-h\alpha_{k-m}$.

The formula (\ref{recur alfa}) defines, by the implicit function theorem, 
the set of quantities $\alpha_k$, $1\le k\le N$, satisfying
\begin{equation}\label{asymp alfa}
\alpha_k=\prod_{j=0}^{m-1}a_{k+j}^{-1}(1+O(h)).
\end{equation}

{\bf Theorem 3.}  {\it The quantities $\alpha_k$ defined by 
{\rm (\ref{recur alfa})} serve as coefficients of the matrix
\begin{equation}\label{Pi+ 3}
L={\cal L}(I+hT^{-m})=I+h\lambda^m\sum_k\alpha_kE_{k+m,k}.
\end{equation}
The discrete time Lax equation
\begin{equation}\label{discrLax 3}
\widetilde{T}=L^{-1}TL={\cal L}^{-1}(I+hT^{-m})\;T\;{\cal L}(I+hT^{-m})
\end{equation}
with the Lax matrix {\rm (\ref{Tmat3})} 
generates the following map on ${\Bbb R}^N\{a\}$, equivalent to 
{\rm (\ref{eq in v 3})}:
\begin{equation}\label{ev 3}
\wa_k=\frac{a_k-h\alpha_{k-m}}{a_{k+m}-h\alpha_k}\;a_{k+m}.
\end{equation}
This map is Poisson with respect to the Poisson bracket {\rm (\ref{Dirac})}
corresponding to the lattice 3, and is interpolated by the flow with the
Hamiltonian function} 
\[
-\frac{1}{m}{\rm tr}\Phi(T^{-m}),\quad where \quad \Phi(\xi)=h^{-1}
\int_0^{\xi}\log(1+h\eta)\frac{d\eta}{\eta}.
\]

{\bf Proof.} Again, it suffices to prove the first two statements.
Assuming for a moment that the ${\cal L}$--factor of the the matrix 
$I+hT^{-m}$ for $T$ from (\ref{Tmat3}) has the form (\ref{Pi+ 3}),
we see that the evolution equation (\ref{discrLax 3}), i.e. 
$L\widetilde{T}=TL$, is equivalent to
\begin{equation}\label{aux 3}
\alpha_k\wa_k=a_{k+m}\alpha_{k+1}, \quad \wa_k+h\alpha_{k-m}=a_k+h\alpha_{k+1}.
\end{equation}
This in turn is equivalent to a combination of an evolution equation
(\ref{ev 3})
with the condition of compatibility of two equations in (\ref{aux 3}):
\begin{equation}\label{comp 3}
a_k-h\alpha_{k-m}=\frac{\alpha_{k+1}}{\alpha_k}(a_{k+m}-h\alpha_k).
\end{equation}
The last equation is equivalent to the fact that
\begin{equation}\label{const 3}
\alpha_k\prod_{j=0}^{m-1}(a_{k+j}-h\alpha_{k+j-m})={\rm const},
\end{equation}
i.e. does not depend on $k$. We shall prove that the actual value of 
this constant is equal to 1, which is equivalent to (\ref{recur alfa}).

To compute the ${\cal L}$--factor of the matrix $I+hT^{-m}$ for $T$ from 
(\ref{Tmat3}), we notice, first, that $T^{-1}=CD^{-1}$, where
\[
C=\lambda\sum_ka_k^{-1}E_{k+1,k},\quad 
D=I+\lambda^{-m}\sum_ka_{k+m}^{-1}E_{k,k+m}.
\]
Further, notice that the ${\cal L}$--factor of any matrix is not changed under
the right multiplication by the factor from ${\rm\bf G}_-$. We multiply the
matrix $I+hT^{-m}=I+(CD^{-1})^m$ from the right by $(DC^{-1})^mC^m$. To see that
this matrix belongs to ${\rm\bf G}_-$, notice that it is equal to 
$DD_1\ldots D_{m-1}$, where 
$D_j=C^{-j}DC^j=I+\lambda^{-m}\sum d_k^{(j)}E_{k,k+m}\in {\rm\bf G}_-$. 
For the further reference we give here an explicit formula
\[
d_k^{(j)}=\prod_{l=0}^{j-1}a_{k+l}\prod_{l=0}^ja_{k+m+l}^{-1}.
\] 
So we get
\[
{\cal L}(I+hT^{-m})={\cal L}(DD_1\ldots D_{m-1} +hC^m),
\] 
and an inspection of this formula convinces that this factor must indeed be of 
the form (\ref{Pi+ 3}), while 
\[
{\cal U}(DD_1\ldots D_{m-1} +hC^m)=\sum_{j=0}^{m}\lambda^{-jm}\sum_k
\beta_k^{(j)}E_{k,k+jm}
\]
Here the quantities  $\alpha_k$, $\beta_k^{(j)}$ 
are completely defined by the recurrent relations following from the 
definitions:
\begin{equation}\label{recur 31}
\alpha_k\beta_k^{(0)}=\prod_{j=0}^{m-1}a_{k+j}^{-1},
\end{equation}
\begin{equation}\label{recur 32}
\beta_k^{(0)}+h\alpha_{k-m}\beta_{k-m}^{(1)}=1,
\end{equation}
\[
\beta_k^{(j)}+h\alpha_{k-m}\beta_{k-m}^{(j+1)}=
{\rm coef.\;\;by}\;\;
\lambda^{-jm}E_{k,k+jm}\;\;{\rm in}\;\;DD_1\ldots D_{m-1},\;\; 1\le j\le m.
\]
(In the last equation for $j=m$ one must set $\beta_k^{(m+1)}=0$, which leads to
$\beta_k^m=\prod_{l=0}^{m-1}a_{k+m^2+l}^{-1}$).

Again, we shall prove that the constant in (\ref{const 3}) is equal to 1, 
only for the two simplest cases $m=1,2$, leaving the calculations for the 
general case to the reader. 

For $m=1$ the defining recurrences (\ref{recur 31}), (\ref{recur 32}) read:
\[
\alpha_k\beta_k^{(0)}=a_k^{-1},\quad
\beta_k^{(0)}+h\alpha_{k-1}\beta_{k-1}^{(1)}=1,\quad
\beta_k^{(1)}=a_{k+1}^{-1}.
\]
Excluding from these relations $\beta_k^{(0)}$, $\beta_k^{(1)}$, we get:
\[
\frac{a_k^{-1}}{\alpha_k}=1-ha_k^{-1}\alpha_{k-1},\quad {\rm or}\quad
\frac{1}{\alpha_k}=a_k-h\alpha_{k-1},
\]
which proves the theorem for $m=1$.

In the case $m=2$ the recurrent relations (\ref{recur 31}), (\ref{recur 32})
take the form
\[
\alpha_k\beta_k^{(0)}=a_k^{-1}a_{k+1}^{-1},\quad
\beta_k^{(0)}+h\alpha_{k-2}\beta_{k-2}^{(1)}=1,
\]
\[
\beta_k^{(1)}+h\alpha_{k-2}\beta_{k-2}^{(2)}=
a_{k+2}^{-1}+a_ka_{k+2}^{-1}a_{k+3}^{-1},\quad
\beta_k^{(2)}=a_{k+4}^{-1}a_{k+5}^{-1}.
\]
Excluding $\beta_k^{(j)}$ from these relations, we get:
\[
\frac{a_k^{-1}a_{k+1}^{-1}}{\alpha_k}=1-h\alpha_{k-2}
(a_k^{-1}+a_{k-2}a_k^{-1}a_{k+1}^{-1}-h\alpha_{k-4}a_k^{-1}a_{k+1}^{-1}),
\]
or
\[
\frac{1}{\alpha_k}=a_{k+1}(a_k-h\alpha_{k-2})-h\alpha_{k-2}
(a_{k-2}-h\alpha_{k-4}).
\]
Using in the last term on the right--hand side (\ref{comp 3}) for $m=2$, 
we can rewrite the last expression as
\[
\frac{1}{\alpha_k}=a_{k+1}(a_k-h\alpha_{k-2})-h\alpha_{k-1}
(a_{k}-h\alpha_{k-2})=(a_{k+1}-h\alpha_{k-1})(a_k-h\alpha_{k-2}).
\]
This finishes the proof for $m=2$. Again, we hope that the pattern of the 
general proof is clear from these two simple cases. It would be highly desirable 
to find a less computational proof for the periodic case of all three lattices.

\section{Conclusion}
A new application of a general scheme for producing integrable discretizations
for integrable Hamiltonian flows is described in the present paper. 
Advantages of this approach are rather obviuos: it is, in principle, applicable
in a standartized way to every system admitting an $r$--matrix formulation,
at least with a constant $r$--matrix satisfying the modified Yang--Baxter
equation. We shall demonstrate elsewhere that the discrete time systems from 
\cite{Discr CM}, \cite{Discr RS} with dynamical $r$--matrices may be also
included into this framework. We hope also to report on numerous further
applications of this approach in the future.

The drawback of this scheme is also obvious to any expert in this field.
Namely, some of the most beautiful discretizations do not live on the same
$r$--matrix orbits as their continuous time counterparts \cite{Ves et al},
\cite{Discr Toda}, \cite{Ragn}, \cite{Discr Garnier}, and there seems to
exist no way of {\it a priori} identifying the correct orbit for nice
discretizations. However, we hope that continuing to collect examples will
someday bring some light to this intriguing problem.

The research of the author is financially supported by the DFG (Deutsche
Forschungsgemeinshaft).


\newpage
\begin{thebibliography}{10}
\bibitem[1]{Ves et al} A.P.Veselov. Integrable systems with discrete time and 
difference operators. -- {\it Funct. Anal. Appl., 1988, v.22, p.1--13};\\
J.Moser, A.P.Veselov. Discrete versions of some classical integrable
systems and factorization of matrix polynomials. -- {\it Commun. Math. Phys., 
1991, v.139, p.217--243};\\
P.Deift, L.-Ch. Li, C.Tomei. Loop groups, discrete versions
of some classical integrable systems, and rank 2 extensions. -- {\it Mem. Amer. 
Math. Soc., 1992, No 479}.

\bibitem[2]{Nij et al} V.Papageorgiou, F.Nijhoff, H.Capel. Integrable mappings 
and nonlinear integrable lattice equations. -- {\it Physics Lett. A, 1990,
v.147, p.106--114};\\
Complete integrability of Lagrangian mappings and lattices of KdV type. 
-- {\it Physics Letters A, 1991, v.155, p.377--387};\\
Integrable time--discrete systems: lattices and mappings -- {\it Lecture 
Notes Math., 1992, v.1510, p.312--325}.

\bibitem[3]{Discr Toda} Yu.B.Suris. Generalized Toda chains in discrete time. -- 
{\it Leningrad Math. J., 1991, v.2,  p.339--352;}\\
Discrete--time generalized Toda lattices: complete integrability and relation 
with relativistic Toda lattices. -- {\it Physics Letters A, 1990, v.145,
p.113--119}.

\bibitem[4]{Ragn} O.Ragnisco. A simple method to generate integrable symplectic
maps -- in: {\it Solitons and chaos, eds. I.Antoniou,
F.J.Lambert (Springer, 1991), p.227--231;}\\
A discrete Neumann system. -- {\it Physics Lett. A, 1992, v.167, 
p.165--171.\\}
Dynamical $r$--matrices for integrable maps. -- {\it Physics Lett. A, 1995,
v.198, p.295--305.}

\bibitem[5]{Discr Garnier} Yu.B.Suris. A discrete--time Garnier system. -- 
{\it Physics Letters A, 1994, v.189, p.281--289;}\\
A family of integrable standard--like maps related to symmetric  spaces. -- 
{\it Physics Letters A, 1994, v.192, p.9--16;}\\
Discrete--time analogs of some nonlinear oscillators in an inverse--square 
potential. --  {\it J. Phys. A: Math. and Gen., 1994, v.27, p.8161--8169.} 

\bibitem[6]{Discr CM} F.Nijhoff, G.-D.Pang. A time discretized version of the 
Calogero--Moser model. -- {\it Physics Lett. A, 1994, v.191, p.101--107;}\\
Discrete--time Calogero--Moser model and lattice KP equation. -- 
{\it In: Proc. Int. Workshop Symmetries and Integrability of difference eqs, 
Eds. D.Levi, L.Vinet, P.Winternitz  (to appear).}

\bibitem[7]{Discr RS} F.Nijhoff, O.Ragnisco, V.Kuznetsov. Integrable 
time--discretization of the Ruijsenaars--Schneider model. -- 
{\it Commun. Math. Phys, 1995 (to appear)}.

\bibitem[8]{New discr Toda} Yu.B.Suris. Bi--Hamiltonian structure of the $qd$ 
algorithm and new discretizations of the Toda lattice. -- {\it Physics 
Letters A, 1995, v.206, p.153--161}.

\bibitem[9]{Discr RTL} Yu.B.Suris. A discrete--time relativistic Toda lattice. -- 
{\it J.Physics A: Math. \& Gen., 1995 (to appear)}.

\bibitem[10]{Discr peakons} Yu.B.Suris. A discrete time peakons lattice. --
{\it Preprint} solv-int 9512001.

\bibitem[11]{B} O.I.Bogoyavlensky. -- {\it USSR Math. Izv., 1987, v.51, and 1988,
v.52}.

\bibitem[12]{rmatBog} Yu.B.Suris. On the $r$--matrix interpretation of Bogoyavlensky
lattices. -- {\it Phys. Letters A, 1994, v.188, p.256--262}.

\bibitem[13]{Hirota} S.Tsujimoto, R.Hirota, S.Oishi. An extension and
discretization of Volterra equation. -- {\it Techn. Report IEICE, NLP 92-90,
1993.}

\bibitem[14]{Itoh} Y.Itoh. Integrals of a Lotka--Volterra system of odd number
of variables. -- {\it Prog. Theor. Phys., 1987, v.78, p.507--510}.

\bibitem[15]{Fuchs} H.Zhang, G.Tu, W.Oevel, B.Fuchssteiner. Symmetries,
conserved quantities, and hierarchies for some lattice systems with soliton
structure. -- {\it J. Math. Phys., 1991, v.33, p.1908--1918}.

\end{thebibliography}
\end{document}